\makeatletter
\declare@file@substitution{revtex4-1.cls}{revtex4-2.cls}
\makeatother

\documentclass[preprint2]{aastex631}

\received{XXX}
\revised{YYY}
\accepted{ZZZ}
\submitjournal{ApJL}

\newcommand{\tbn}{$\theta_{Bn}$}

\newcommand{\noopsort}[1]{}

\begin{document}
\title{Irregular proton injection to high energies at interplanetary shocks}

\correspondingauthor{Domenico Trotta}
\email{d.trotta@imperial.ac.uk}

\author[0000-0002-0608-8897]{Domenico Trotta}
\affiliation{The Blackett Laboratory, Department of Physics, Imperial College London, London SW7 2AZ, UK}

\author[0000-0002-7572-4690]{Timothy S. Horbury}
\affiliation{The Blackett Laboratory, Department of Physics, Imperial College London, London SW7 2AZ, UK}

\author[0000-0002-3176-8704]{David Lario}
\affiliation{Heliophysics Science Division, NASA Goddard Space Flight Center, Greenbelt, MD 20771, USA}

\author[0000-0002-0074-4048]{Rami Vainio}
\affiliation{Department of Physics and Astronomy, University of Turku, Finland}

\author[0000-0003-3903-4649]{Nina Dresing}
\affiliation{Department of Physics and Astronomy, University of Turku, Finland}

\author[0000-0003-1589-6711]{Andrew Dimmock}
\affiliation{Swedish Institute of Space Physics, Uppsala, Sweden}

\author[0000-0002-0850-4233]{Joe Giacalone}
\affiliation{Lunar and Planetary Laboratory, University of Arizona, Tucson, USA}

\author[0000-0002-3039-1255]{Heli Hietala}
\affiliation{School of Physics and Astronomy, Queen Mary University of London, London E1 4NS, UK}

\author[0000-0002-7388-173X]{Robert F. Wimmer-Schweingruber}
\affiliation{Institute of Experimental and Applied Physics, Kiel University, 24118 Kiel, Germany}

\author[0000-0001-7846-804X]{Lars Berger}
\affiliation{Institute of Experimental and Applied Physics, Kiel University, 24118 Kiel, Germany}

\author[0000-0002-6416-1538]{Liu Yang}
\affiliation{Institute of Experimental and Applied Physics, Kiel University, 24118 Kiel, Germany}



\begin{abstract}
How thermal particles are accelerated to suprathermal energies is an unsolved issue, crucial for many astrophysical systems. We report novel observations of irregular, dispersive enhancements of the suprathermal particle population upstream of a high-Mach number interplanetary shock. We interpret the observed behavior as irregular ``injections'' of suprathermal particles resulting from shock front irregularities. Our findings, directly compared to self-consistent simulation results, provide important insights for the study of remote astrophysical systems where shock structuring is often neglected.
\end{abstract}

\keywords{Acceleration of particles --- plasmas -- shock waves --- Sun: heliosphere --- Sun: solar wind}


\section{Introduction} \label{sec:intro}
Collisionless shock waves are fundamental sources of energetic particles, which are ubiquitously present in our universe and pivotal to explain many of its features, such as the non-thermal radiation emission common to many astrophysical sources, as revealed by decades of remote and direct observations \citep{Reames1999,Amato2017}. Particle acceleration to suprathermal energies from thermal plasma, less understood than particle acceleration starting from an already energised population, remains a puzzle, and has been object of extensive theoretical and numerical investigations~\citep{Drury1983, Caprioli2014, Trotta2021}. 

Shocks in the heliosphere, unique as directly accessible by spacecraft~\citep{Richter1985}, provide the missing link to remote observations of astrophysical systems. Direct observations of the Earth's bow shock using single and multi-spacecraft approaches~\citep[e.g.,][]{Johlander2016} reveal a complex scenario of energy conversion and particle acceleration at the shock transition~\citep{Amano2020, Schwartz2022}. The emerging picture, well supported by theory and modelling, is that small scale irregularities in the spatial and temporal evolution of the shock environment~\citep[][]{Greensadt1980, Matsumoto2015} are fundamental for efficient ion injection to high energies~\citep{Dimmock2019}. This idea of irregular particle injection has been investigated in the past for the Earth's bow shock~\citep{Madanian2021} and in numerical simulations~\citep{Guo2013}, thus suggesting that particle behaviour at shocks is much more complex than what is expected neglecting space-time irregularities, as suggested by early theoretical and numerical works~\citep{Decker1990,Ao2008,Lu2009}.
 
Such a complex picture is not as well observed and understood for shocks beyond the Earth's bow shock. In particular, shock structuring at Interplanetary (IP) shocks, generated as a consequence of phenomena such as Coronal Mass Ejections~\citep[CMEs,][]{Gosling1974} and its role in particle acceleration remains elusive~\citep{BlancoCano2016,Kajdic2019}. IP shocks are generally weaker and have larger radii of curvature with respect to Earth's bow shock, allowing for direct observations of collisionless shocks in profoundly different regimes~\citep[e.g.,][]{Kilpua2015, Yang2020}, and are more relevant to astrophysical environments such as galaxy cluster shocks, where shock irregularities are not resolved, but they are likely to play a crucial role in efficient particle acceleration~\citep[][]{Brunetti2014}. Therefore, the study of particle injection at IP shocks is fundamental to test our current understanding built on Earth's bow shock, as well for addressing shocks at objects currently beyond reach. This paper demonstrates that, in order to address the suprathermal particle production upstream of supercritical collisionless shocks, the inherent variability of the injection process in both time and space must be taken into account.

The Solar Orbiter mission \citep[SolO,][]{Muller2020} probes the inner heliosphere with unprecedented levels of time-energy resolution for energetic particles, thus opening a new observational window for particle acceleration. In this work, we study the acceleration of low-energy ($\sim$ 1 keV) particles to supra-thermal energies ($\sim$ 50 keV) at a strong IP shock observed by SolO at heliocentric distance of about 0.8 AU on 2021 October 30$^{\mathrm {th}}$ at 22:02:07 UT. We use the SupraThermal Electrons and Protons sensor (STEP) of the Energetic Particle Detector (EPD) suite \citep{RodriguezPacheco2020}, measuring particles in the 6 - 60 keV energy range (close to the injection range), at the very high time resolution of 1 s, close to suprathermal particle gyroscales. Our work exploits such novel, previously unavailable datasets for suprathermal particles upstream of IP shocks. We resolve upstream enhancements in the suprathermal particle population with dispersive velocity signatures, and link them to irregular proton injection along the shock front. Our findings are corroborated by kinetic simulations showing similar irregular proton energization upstream close to the shock, thus elucidating the mechanisms responsible for this behaviour. This letter is organised as follows: results are presented in Section~\ref{sec:results}. SolO observations are shown and discussed in Section~\ref{subsec:observations}, while modelling results are reported in~\ref{subsec:modelling}. The conclusions are in Section~\ref{sec:conclusions}.

\section{Results} \label{sec:results}
\subsection{Solar Orbiter Observations}\label{subsec:observations}

%+++++++Figure 1++++++++++++++++
\begin{figure}
\includegraphics[width=.5\textwidth]{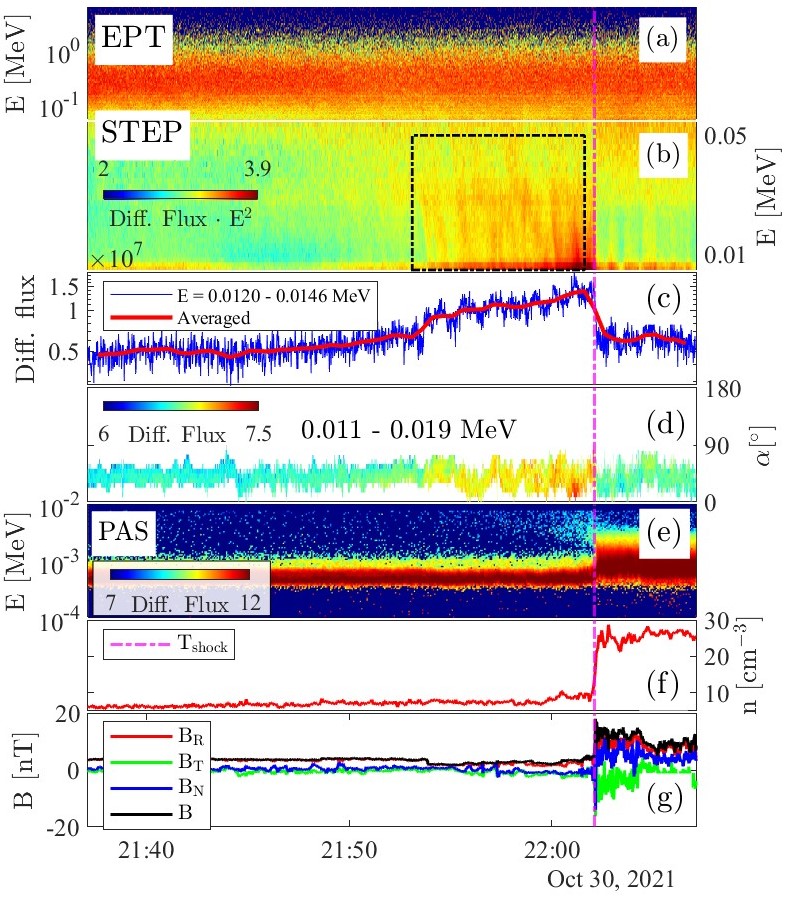}
\caption{Event overview. (a) EPD-Electron Proton Telescope (EPT) particle flux (sunward aperture). (b) EPD-STEP particle flux (magnet channel averaged over the entire field of view). (c) Pitch angle distributions for ions with an energy of 0.011 - 0.019 MeV in the spacecraft frame. (d) Time profile of the STEP energy flux in the 0.012 - 0.015 MeV energy channel at full resolution (blue), and time-averaged using a 1 minute window. (e) SWA-PAS ion energy flux~\citep{Owen2020}. (f) SWA-PAS proton density. (g) MAG burst magnetic field data in RTN coordinates~\citep{Horbury2020}. The magenta line marks the shock crossing, and the black rectangle highlights the dispersive energetic particle enhancements observed by STEP. Differential fluxes are in $\mathrm{E^2 \cdot cm^{-2} s^{-1} sr^{-1} MeV}$ for the EPD instruments and $\mathrm{cm^{-2} s^{-1} eV}$ for PAS. }
\label{fig:fig1_overview}
\end{figure}
%+++++++++++++++++++++++++++++++

Fig.~\ref{fig:fig1_overview} shows a 30 minute overview across the shock transition. Panels (a)-(b) reveal the presence of shock accelerated particles at energies of up to 100 keV, while particle fluxes at higher energies do not respond to the shock passage. At these high energies the fluxes were enhanced following a large Solar Energetic Particle (SEP) event \citep[see][]{Klein2022}.

The most striking feature of the period prior to the shock arrival at SolO is the irregular energetic particle enhancements particularly evident at 10 - 30 keV energies (Fig.~\ref{fig:fig1_overview} (b), black box), found in the time interval $~\sim$ 15 minutes before the shock crossing, corresponding to $ 2\times 10^{5}$ km or 2500 ion inertial lengths, $d_i$. These particle enhancements have the novel feature of being dispersive in energy and are the focus of this work. The typical timescales at which the irregularities are observed are of 10-20 seconds, corresponding to spatial scales of about 50 $d_i$.  Such signatures were previously inaccessible to observations, as shown in Fig.~\ref{fig:fig1_overview} (c), where the time profile of ion differential flux in the 0.012 - 0.015 MeV channel, rising exponentially up to the shock ~\citep{Giacalone2012}, is shown at full resolution (blue) and averaged using a $\sim$ 1 minute window, typical of previous IP shock measurements. Fig.~\ref{fig:fig1_overview}(d) shows pitch angle intensities for 0.011 -- 0.019 MeV ions (i.e., energies at which the irregular enhancements are observed). Pitch angles are computed in the plasma rest frame assuming that all ions are protons, and performing a Compton-Getting correction~\citep{ComptonGetting1935}, thereby combining magnetic field data from the magnetometer~\citep[MAG,][]{Horbury2020}, and solar wind plasma data from the Proton and Alpha particle Sensor (PAS) on the Solar Wind Analyser (SWA) instrument suite~\citep{Owen2020}, and particle data from EPD/STEP~\citep{Yang2023}. For the interval studied, low pitch angles are in the 30$^\circ$ field of view of STEP, relevant for shock reflected particles. The irregular enhancements of energetic particles are field aligned, as is evident for the strongest signal close to the shock transition. The flux enhancement visible in PAS (Fig.~\ref{fig:fig1_overview}(e)) at lower energies starting immediately before the shock (22:00 UT) also reveals a field-aligned population. The study of the PAS low-energy population and the behaviour very close to the shock transition is object of another investigation~\citep{Dimmock2023}.

%+++++++Figure 2 -- Dispersive signal fit++++++++++++++++
\begin{figure*}
	\includegraphics[width=\textwidth]{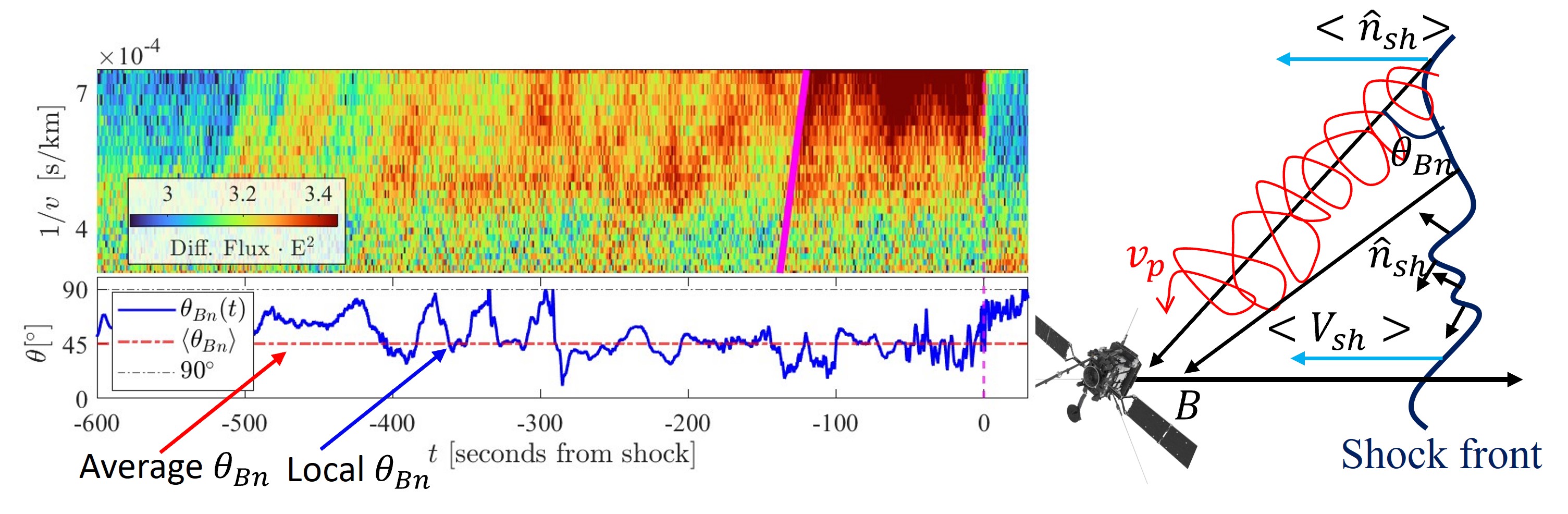}
	\caption{\textit{Left}: Spectrogram of the irregular signal in seconds from shock vs $1/v$ axes, with the velocity dispersion shown by the solid magenta line (top).  Time series showing the local $\theta_{Bn}(t)$ angle. The red and grey dashed lines represent the average $\theta_{Bn}$ and a 90$^\circ$ angle, respectively (bottom). \textit{Right}: Cartoon showing the corrugated shock front with local shock normal, trajectory of a reflected particle and the Solar Orbiter trajectory (SolO model: esa.com).}
	\label{fig:fig2_fit}
\end{figure*}
%+++++++++++++++++++++++++++++++

%	Further analyses of the behaviour of suprathermal protons in pitch angle space will be object of future studies looking at more events.}

 The magnetic field reveals a wave foreshock $\sim$ 2 minutes upstream of the shock, in conjunction with a population of low-energy ($\sim$ 4 keV) reflected particles seen by SWA/PAS, visible as the light blue enhancement in Fig.~\ref{fig:fig1_overview}(e) around 22:00 UT. Interestingly, the magnetic field is quieter where signals of irregular injection are found, indicating that efficient particle scattering may be reduced in this region~\citep{Lario2022}. In this ``quiet'' shock upstream, we found two structures compatible with shocklets in the process of steepening ($\sim$21:57 UT), very rarely observed at IP shocks \citep{Wilson2009, Trotta2023a}.

The shock parameters were estimated using upstream/downstream averaging windows varied systematically between 1 and 8 minutes~\citep{Trotta2022b}. The shock was oblique, with a normal angle $\theta_{Bn} = 44 \pm 1.5^\circ$ (obtained with the Mixed Mode 3 technique~\citep[MX3][]{Paschmann2000}, compatible with MX1,2 and Magnetic Coplanarity). The shock speed in the spacecraft frame and along the shock normal is $\mathrm{V_{shock} = 400 \pm 5\, km/s}$. The shock Alfv\'enic and fast magnetosonic Mach numbers are $\rm M_A \sim 7.6$ and $\rm M_{fms} \sim 4.6$, respectively. Thus, the event provides us with the opportunity to study a shock with particularly high Mach number in comparison with other IP shocks, while the shock speed is moderate with respect to typical IP shocks~\citep{Kilpua2015}. The shock is supercritical, and therefore expected to have a corrugated, rippled front~\citep{Trotta2019, Kajdic2021JGR}. The presence of reflected particles, enhanced wave activity in close proximity (1 minute) to the shock transition  and upstream shocklets in the process of steepening is consistent with the local shock parameters~\citep{BlancoCano2016}.

To further elucidate the dispersive nature of the suprathermal particles, we show the STEP energy spectrogram in $1/v$ vs $t$ space (Fig.~\ref{fig:fig2_fit}). Here, particle speeds are referred to the center of the relative energy bin and computed in the spacecraft rest frame, assuming that all particles detected are protons~\citep[see][ for further details]{Wimmer2021}. During the period of irregular particle enhancements, we also combined magnetic field and plasma data to compute the particle pitch angles in the solar wind frame~\citep{Compton1935}, revealing that the particles detected by STEP are closely aligned with the field (not shown here). Interestingly, by visual inspection, it can be seen that these dispersive signals are shallower going far upstream, consistent with the fact that they are injected from more distant regions of the shock.

The dispersive flux enhancements are associated with irregular acceleration of protons along the shock front. Indeed, due to their dispersive nature, the particles detected by STEP cannot be continuously produced at the shock and propagated upstream, but they must come from a source that is only temporarily magnetically connected to the spacecraft due to time and/or space irregularities. Then, the fastest particles produced at the irregular source are detected first by the spacecraft, followed by the slower ones, yielding the observed dispersive behaviour. Given the short timescales at which energetic particle enhancements are observed with respect to the shock and the quiet behaviour of upstream magnetic field in the 10 minutes upstream of the shock, we assume that particles do not undergo significant scattering from their (irregular) production to the detection at SolO. 
%However, it is important to note that the observed particles, will be scattered by pre-existing and/or self-generated turbulence,~\citep{Kis2007} relevant for particle acceleration mechanisms further accelerating particles to higher energies~\citep{Drury1983}. 
It is then natural to investigate the connection with the shock. The bottom-left panel of Fig.~\ref{fig:fig2_fit} shows the local $\theta_{Bn}(t) \equiv \cos^{-1}\left( \mathbf{B}(t) \cdot \hat{\mathbf{n}}_{\rm shock} / |\mathbf{B}(t)| \right)$ changing significantly when the dispersive signals are observed, indicating that the spacecraft was indeed connected to different portions of the (corrugated) shock front, which in turn is expected to respond rapidly to upstream changes, as recent simulation work elucidated~\citep[e.g.,][]{Trotta2023b}. Note that, given the single-spacecraft nature of the observations, the average shock normal computed with MX3 for both local and average \tbn estimation was used.

To further support this idea, similarly to Velocity Dispersion Analyses (VDA) used to determine the injection time of SEP events~\citep[e.g.,][]{Lintunen2004, Dresing2023}, we chose the clearest dispersive signal ($\sim$ 100 seconds upstream of the shock) and we superimpose the following relation (indicated by the magenta line in Fig.~\ref{fig:fig2_fit}):
\begin{equation}
    \label{eq:vda}
    t_{\mathrm{O}} (v) = t_i + \frac{s}{v},
\end{equation}
where $t_\mathrm{O}$ represents the time at which the flux enhancement is observed for a certain speed $v$, $t_i$ is the time of injection at the source, and $s$ is the distance travelled by the particles from the source to the spacecraft. Thus, the argument is that the dispersive signals are due to accelerated particles produced by different portions of the shock front temporarily connected with the spacecraft, as sketched in Fig.~\ref{fig:fig2_fit} (right). We note that, due to the very high energy-time resolution of STEP, it was possible to perform the VDA on such small ($\sim$seconds) time scales. Determining $t_i$ based on the time when the highest energy particles are observed ($t_i \sim -130 s$), the source distance that we obtain through Equation~\ref{eq:vda} is $s \approx 4 \times 10^4$ km ($\sim 500 d_i$), compatible with their generation at the approaching shock, for which we would expect $s \sim \mathrm{V_{shock} \Delta t / sin(\theta_{Bn})}$, where $\mathrm{V_{shock}}$ is the average shock speed, $\Delta t$ is the time delay between the observation of the dispersive signal and the shock passage. This is also compatible with the fact that the other dispersive signals observed further upstream, such as the one before 21:54, about 500 seconds upstream of the shock (see Fig.~\ref{fig:fig2_fit}), show a shallower inclination, though a more precise, quantitative analysis of this behaviour is complicated by the high noise levels of the observation, and will be the object of later statistical investigation employing more shock candidates~\citep{Yang2023}.

%+++++++Figure 5 -- Simulations++++++++++++++++
\begin{figure}
\includegraphics[width=.45\textwidth]{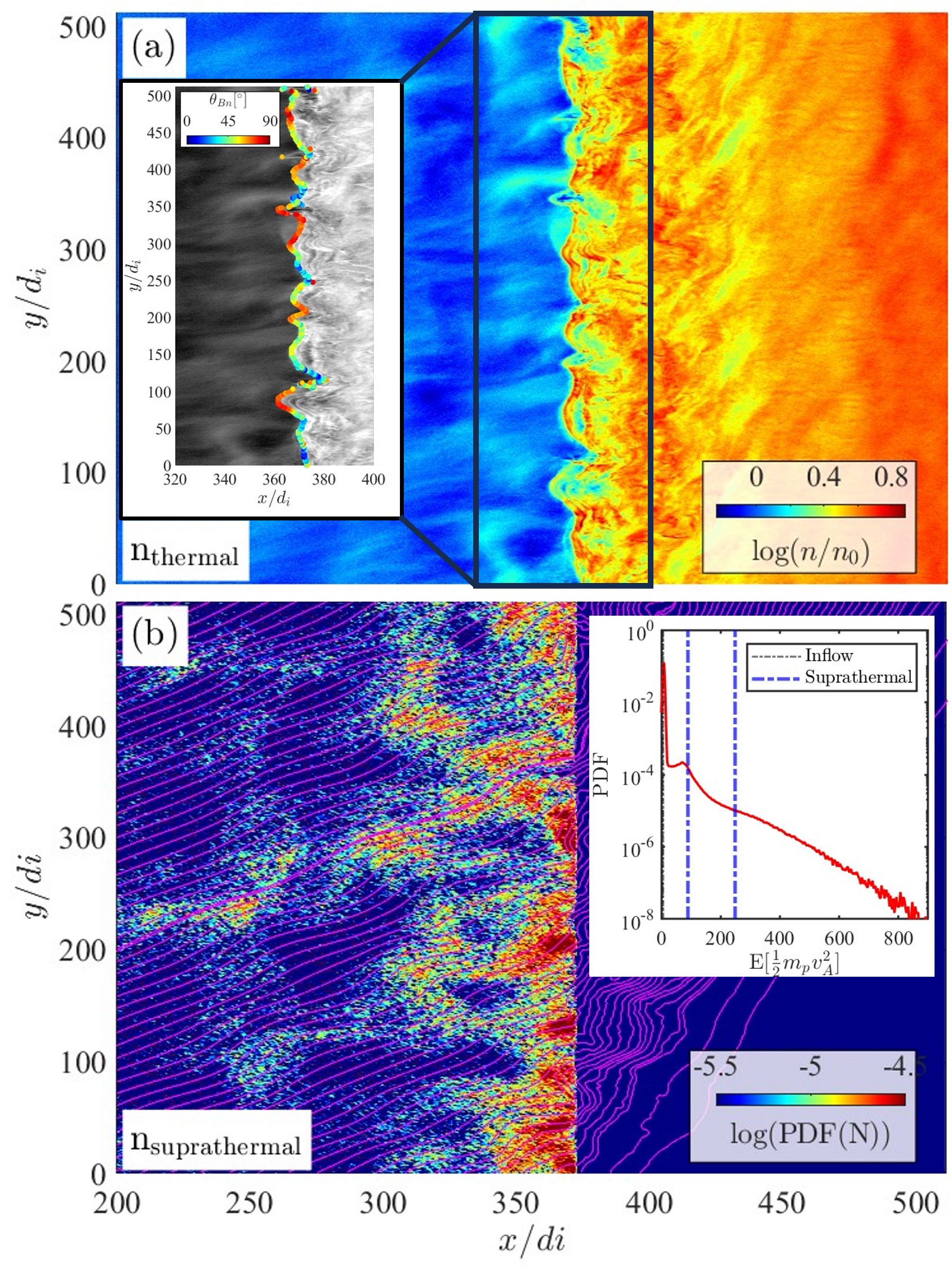}
\caption{\textit{Top}: Simulation snapshot of proton density (colormap). The inset shows a zoom around the shock transition (grey), and the local shock position is superimposed, with a colormap correesponding to the local \tbn. \textit{Bottom}: Density map of upstream superathermal protons (colormap) and magnetic field lines (magenta) computed at the same simulation time as (a). The inset shows the upstream particle energy spectrum, with the dashed blue lines indicating the suprathermal energy range considered.}
\label{fig:fig4_sim}
\end{figure}
%+++++++++++++++++++++++++++++++

\subsection{Shock Modelling}\label{subsec:modelling}
Further insights about shock front irregularities are limited by the single-spacecraft nature of these observations. Therefore, we employ 2.5-dimensional kinetic simulations, with parameters compatible with the observed ones, to model the details of the shock transition, where proton injection to suprathermal energies takes place, relevant to our interpretation of the dispersive signals and enabling us to see how the shock surface and normal behave at small scales (see Fig.~\ref{fig:fig2_fit}). In the simulations, protons are modelled as macroparticles and advanced with the Particle-In-Cell (PIC) method, while the electrons are modelled as a massless, charge-neutralizing fluid~\citep{Trotta2020a}.

In the model, distances are normalised to the ion inertial length $d_i$, times to the upstream inverse cyclotron frequency ${\Omega_{ci}}^{-1}$, velocity to the Alfv\'en speed $v_A$, and the magnetic field and density to their upstream values $B_0$ and $n_0$. The shock is launched with the injection method~\citep{Quest1985}, where an upstream flow speed $V_\mathrm{in} = 4.5 v_A$ was chosen, corresponding to $M_A \sim 6$. The shock nominal $\theta_{Bn}$ is 45$^\circ$. The simulation domain is 512 $d_i$ $\times$ 512 $d_i$, with resolution $\Delta x$ = $\Delta y$ = 0.5 $d_i$ and a particle time-step $\Delta t_{pa}$  = 0.01 $\Omega_{ci}^{-1}$. The number of particles per cell used is always greater than 300. This choice of parameters is compatible with the local properties of the IP shock as estimated from the SolO measurements. However, inherent variability routinely found in the simulations at small scales and in the observations at larger scales must be considered when comparing numerical and observational results. We note that these simulations are initialised with a laminar upstream, and therefore the fluctuations that impact the shock are self-generated (due to particle reflection and subsequent upstream propagation). An exhaustive characterization of these self-induced fluctuations is discussed in \citet{Kajdic2021JGR}.

Simulation results are shown in Fig.~\ref{fig:fig4_sim}. In the top panel, we present the proton density for a simulation snapshot where the shock transition is well-developed, showing the strongly perturbed character of the shock front. In such an irregular shock transition, particle dynamics become extremely complex~\citep[e.g.,][]{Lembege1992}. To further elucidate the irregularities of the shock front, we computed the shock position in the simulation domain (with the criterion $B>3B_0$, as in \citet{Trotta2023b}) and evaluated the local $\theta_{Bn}$ along it (Fig.~\ref{fig:fig4_sim}(a), inset), showing high variability (see the sketch in Fig.~\ref{fig:fig2_fit}).

In the bottom panel of Fig.~\ref{fig:fig4_sim}, we study the self-consistently shock-accelerated protons. The upstream energy spectrum is shown in the inset, with a peak at the inflow population energies and a suprathermal tail due to the accelerated protons. To address particle injection, we analyse the upstream spatial distribution of such suprathermal protons (Fig.~\ref{fig:fig4_sim}(b)) at the energies highlighted in the inset, which are a factor of 10 larger than the typical energies of particles in the upstream inflow population, in a similar fashion as the energy separation between the STEP energies at which the irregular enhancements are observed ($\sim$ 10 keV) and the Solar wind population energies measured by PAS ($\sim$ 1 keV) . It can be seen that suprathermal particles are not distributed uniformly, and their spatial distribution varies with their locations along the shock front, another indication of irregular injection. Furthermore, we observed that the length scale of the irregularities is of 50 $d_i$, directly comparable with the irregularities seen in the STEP fluxes (see Fig.~\ref{fig:fig1_overview}). Higher energy particles also show irregularities.

\section{Conclusions} \label{sec:conclusions}
We studied irregular particle acceleration from the thermal plasma using novel SolO observations. Particle injection to high energies is an extremely important issue for a large collection of astrophysical systems making the SolO shock on 2021 October 30$^{\mathrm{th}}$ an excellent event to tackle this interesting problem. The capabilities of the SolO EPD suite were exploited to probe the complex shock front behaviour in the poorly investigated IP shock case. From this point of view, \textit{in-situ} observations of irregular particle enhancements have been used as a tool to address the (remote) structuring of the shock, an information not available by simply looking at the spacecraft shock crossing of in one point in space and time. Such an approach is reminiscent to the ones used to reconstruct the properties of SEP events~\citep{Krucker1999}, and even to the ones looking at the properties of the heliospheric termination shock with the Interstellar Boundary Explorer mission \citep[IBEX,][]{McComas2009}, where particles produced at different portions of the shock are used to understand its dynamics~\citep{Zirnstein2022}.

%The shock was oblique, with high Alfve'nic and fast magnetosonic Mach numbers, and thus supercritical and expected to show irregularities (corrugation, rippling). We could resolve upstream, dispersive enhancements in suprathermal particle fluxes, which we linked to irregular proton acceleration happening at different portions of the corrugated shock front, where the spacecraft was temporarily connected to.  Thus, 
The hybrid kinetic simulations are consistent with this complex scenario of proton acceleration, with irregularly distributed suprathermal particles along the shock front, an invaluable tool to elucidate the small-scale behaviour of this IP shock and of shock transitions in a variety of astrophysical systems. Our model highlights the very small-scale behaviour of the shock, but neglects other effects like pre-existing turbulence and interplanetary disturbances that may be important~\citep{Lario2002, Trotta2022a, Nakanotani2022, Trotta2023b}. The direct investigation of shock acceleration in systems other than the Earth's bow shock (having a small radius of curvature and many other properties important for planetary bow shocks) is important to build a comprehensive understanding of collisionless shocks energetics. This work significantly strengthens an evolving theory of collisionless shock acceleration. Combining high resolution energetic particle data upstream of heliospheric shocks with hybrid simulations, we have shown, for interplanetary shocks, that the inherent variability of the injection process in both time and space must be considered to solve the problem of how suprathermal particle injection occurs in astrophysical systems.  The process analysed here is general, as it does not depend on how shock irregularities are generated. Indeed, this study is relevant for astrophysical systems where shock front irregularities cannot be resolved but are likely to play an important role for particle acceleration from the thermal distribution, such as galaxy cluster shocks, where efficient particle acceleration, which is inferred to happen at very large, $\sim$ Mpc scales, remains a puzzle, particularly in the absence of pre-existing cosmic rays~\citep{Botteon2020}.

%% Also note that the akcnowlodgment environment does not support long amounts of text. If you have a lot of people and institutions to acknowledge, do not use this command. Instead, create a new \section{Acknowledgments}.
\begin{acknowledgments}
This study has received funding from the European Unions Horizon 2020 research and innovation programme under grant agreement No. 101004159 (SERPENTINE, www.serpentine-h2020.eu). Part of this work was performed using the DiRAC Data Intensive service at Leicester, operated by the University of Leicester IT Services, which forms part of the STFC DiRAC HPC Facility (www.dirac.ac.uk), under the project ``dp031 Turbulence, Shocks and Dissipation in Space Plasmas''. N.D. acknowledges the support of the Academy of Finland (SHOCKSEE, grant nr. 346902). H.H. is supported by the Royal
Society University Research Fellowship URF\textbackslash R1\textbackslash 180671. D.L. acknowledges support from NASA Living With a Star (LWS) program NNH19ZDA001N-LWS, and the Goddard Space Flight Center Heliophysics Innovation Fund (HIF) program.
\end{acknowledgments}

\end{document}